\documentclass[12pt,preprint]{aastex}
\pdfoutput=1

\usepackage{color}

\title{Detecting a Unique EBL Signature with TeV Gamma Rays}
\author{A. Imran$^{1}$, F. Krennrich$^{1}$.}
\affil{$^1$ Dept. of Physics \& Astronomy, Iowa State University, Ames, IA, 50011, USA}

\email{imranisu@iastate.edu}

\begin{abstract}We discuss prospects for detecting a spectral break in $\gamma$-ray 
spectra of blazars due to the extragalactic background light (EBL) density
falling off between the near and mid-IR.  A measurable spectral change 
in the TeV spectra at 1~TeV could arise from a rapid or slow drop in the EBL 
density above $\approx$~1~um. This effect is mediated by the ratio of the 
near to mid-IR density of EBL. A detection  of such a spectral feature 
could become a clear signature of EBL absorption. A non-detection would give 
a strong observational constraint to the shape of the EBL spectrum.  We present 
calculations estimating the sensitivity of TeV telescopes for detecting such a 
break for blazar observations at different redshifts.
\end{abstract}


\begin{document}
\maketitle
%Begin the section.

 %=======================================
\section{Introduction}
 %=======================================
Very high energy $\gamma$-ray beams from sources at cosmological distances provide a unique opportunity to  probe the intervening medium.  The interaction is with photons from the extragalactic background light (\cite{gs67}, \cite{sds92}) via $\gamma+\gamma \rightarrow e^+ + e^-$.
 Direct measurements of the EBL, particularly  in the mid-IR, are extremely  difficult due to strong  foreground emissions, making TeV $\gamma$-ray absorption measurements invaluable. 

The task of extracting information about the EBL from spectra of TeV $\gamma$-ray sources is complicated by our lack of knowledge about the intrinsic source spectra. Attempts to find signatures from the EBL in TeV spectra \cite{sds92} were hindered by the possibility that cutoffs could be intrinsic. As a result, current $\gamma$-ray based methods to probe the diffuse background (\cite{dk2005}, \cite{aha2006}) necessarily rely on assumptions made about the source spectra and are often founded on a theoretical understanding of particle acceleration and emission mechanisms in blazar jets. An unambigous EBL feature independent from source properties is highly desirable for getting a better understanding of the level of $\gamma$-ray absorption on extragalactic distance scales.  
A new generation of highly sensitive air Cherenkov telescopes (HESS, MAGIC and VERITAS) combined with strong blazar flares might allow the detection of a clear signature from the attenuation of TeV photon by the EBL. We discuss the relation between the relative intensity of the EBL in the near and mid-IR range and how it affects the rise in TeV opacity with increasing energy. Consequently, a change in opacity could alter the TeV spectrum around 1 TeV, potentially leading to a detectable spectral hardening or softening that is unique to the shape and intensity of the EBL. To characterize the affect of the relative EBL intensity in the near and mid-IR regime, a number of viable EBL scenarios \cite{dk2005} are applied to a set of simulated blazar spectra over a range of redshifts. Specifically, we explore the sensitivity required to detect such an EBL imprint in the TeV spectra of typical blazar flaring levels with the HESS or VERITAS array.

 %=======================================
\section{Spectral break for different EBL scenarios}
 %=======================================
 The spectral energy distribution of the EBL in the optical to far-IR light is generally thought to exhibit a bimodal shape (Fig.1). The consecutive peaks at $\sim$~1 and $\sim$~100 $\mu$m are attributed to starlight and  dust emission \cite{hd98}. Following the approach by \cite{dk2005}, we utilize  12 different EBL realizations consistent with limits from direct EBL measurements\footnote{These EBL scenarios are consistent with  observational limits in the near, mid-IR and far-IR}. The HLL scenario stands for a 'High' near-IR, 'Low' mid-IR and 'Low' far-IR (Fig.1).  In Fig. 2 we show the opacity versus energy for some EBL scenarios: whereas HLL shows a peak at 1 TeV and a drop up to 10 TeV, the  LHH case shows a steeply rising EBL at those energies.  The opacity function for the HLL scenario translates into spectral hardening, for the LHH it causes spectral softening in aborbed $\gamma$-ray spectra.  Most importantly, both spectral features occur at $\approx$~1~TeV, corresponding to the density peak in the near IR at $\rm 1\mu m$, hence providing a clear search strategy.  

\begin{figure}
\begin{center}
\includegraphics [scale=0.32, angle=0]{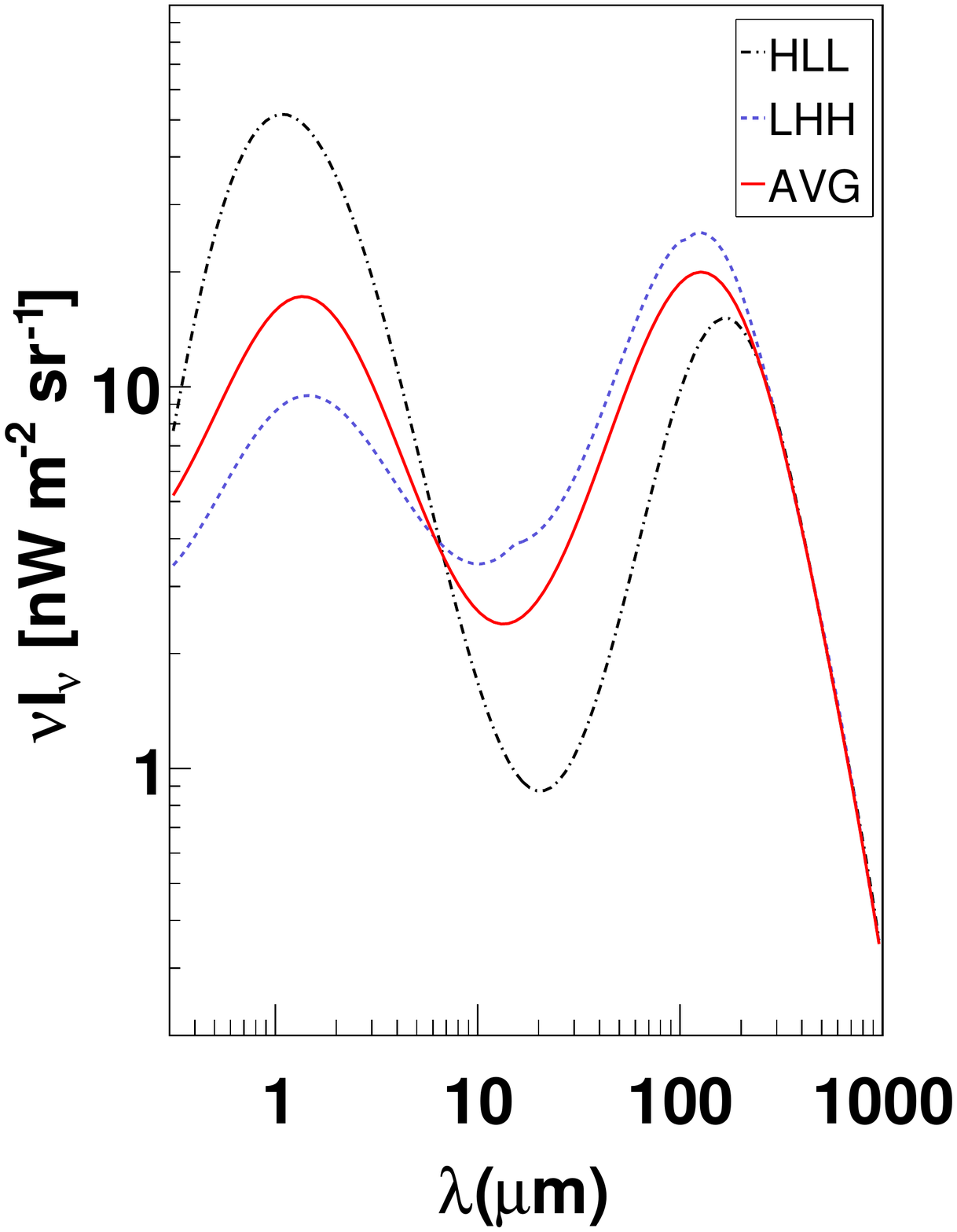}
\end{center}
\caption{Template spectra of EBL realizations with substantially different near- to mid-IR
              density ratios are shown.}\label{fig1}

\end{figure}
\begin{figure}
\begin{center}
\includegraphics [scale=0.32, angle=0]{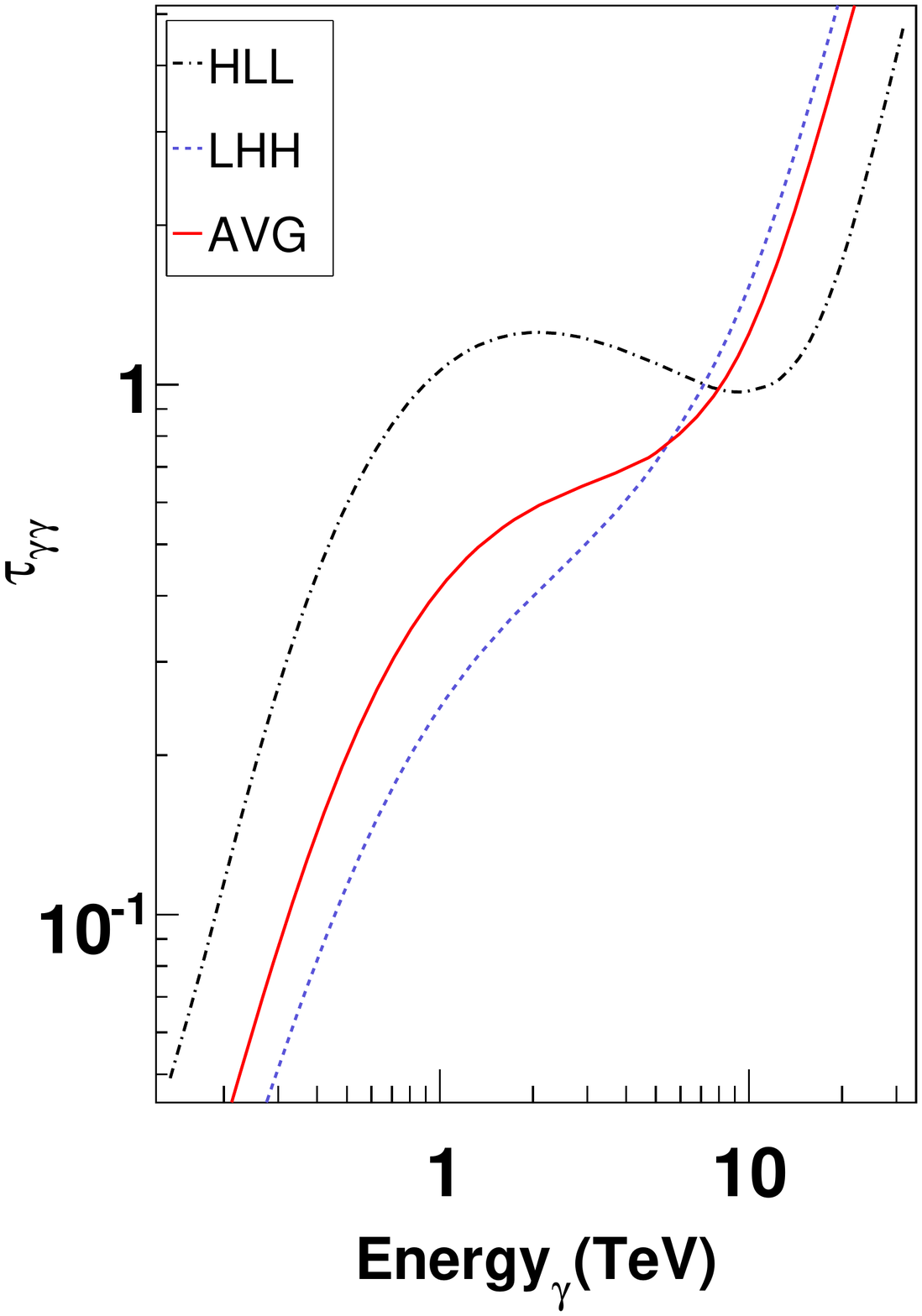}
\end{center}
\caption{The opacity due to 3 distinct EBL scenarios for a source at {\it z}=0.03 is shown
                   as a function of energy. }\label{fig2}
\end{figure}

TeV blazars exhibit flares over time scales of weeks, allowing exposure times of $\ge$~50 hours with Cherenkov telescopes. Recorded flux levels historically range between $\approx$~0.1 - 10 Crab. In fact Mrk 421, Mrk 501 ({\it z} =0.03) have shown high flaring levels of $\ge$ 4 Crab for longer than several weeks.  When correcting for distance, many blazars show similar intrinsic flaring levels: 1ES1959+650, PKS2155-304\footnote{The recent report of  a flare by the HESS collaboration with PKS2155 reaching 15 Crab at the objects true redshift of 0.115 make our assumptions conservative.}, H1426+428 and 1ES1101-232.  To evaluate the prospects of finding an EBL induced absorption signature in the TeV blazars, we construct a set of hyothetical blazar spectra with  power law spectral indices of $\alpha$ =-1.6 to -2.4,  placed at redshifts of {\it z} =0.03 to 0.186. These model blazars are normalized to a flux level of $\sim$ 4 Crab at 200 GeV and a reshift of 0.03. The instrument response with typical statistical uncertainties associated with such spectra are derived, e.g., from the HESS detection of 1ES1101-232 \cite{aha2006} and are also folded in to our test blazars.

\begin{figure}
\begin{center}
\includegraphics [scale=0.75 , angle=90,width=0.48\textwidth]{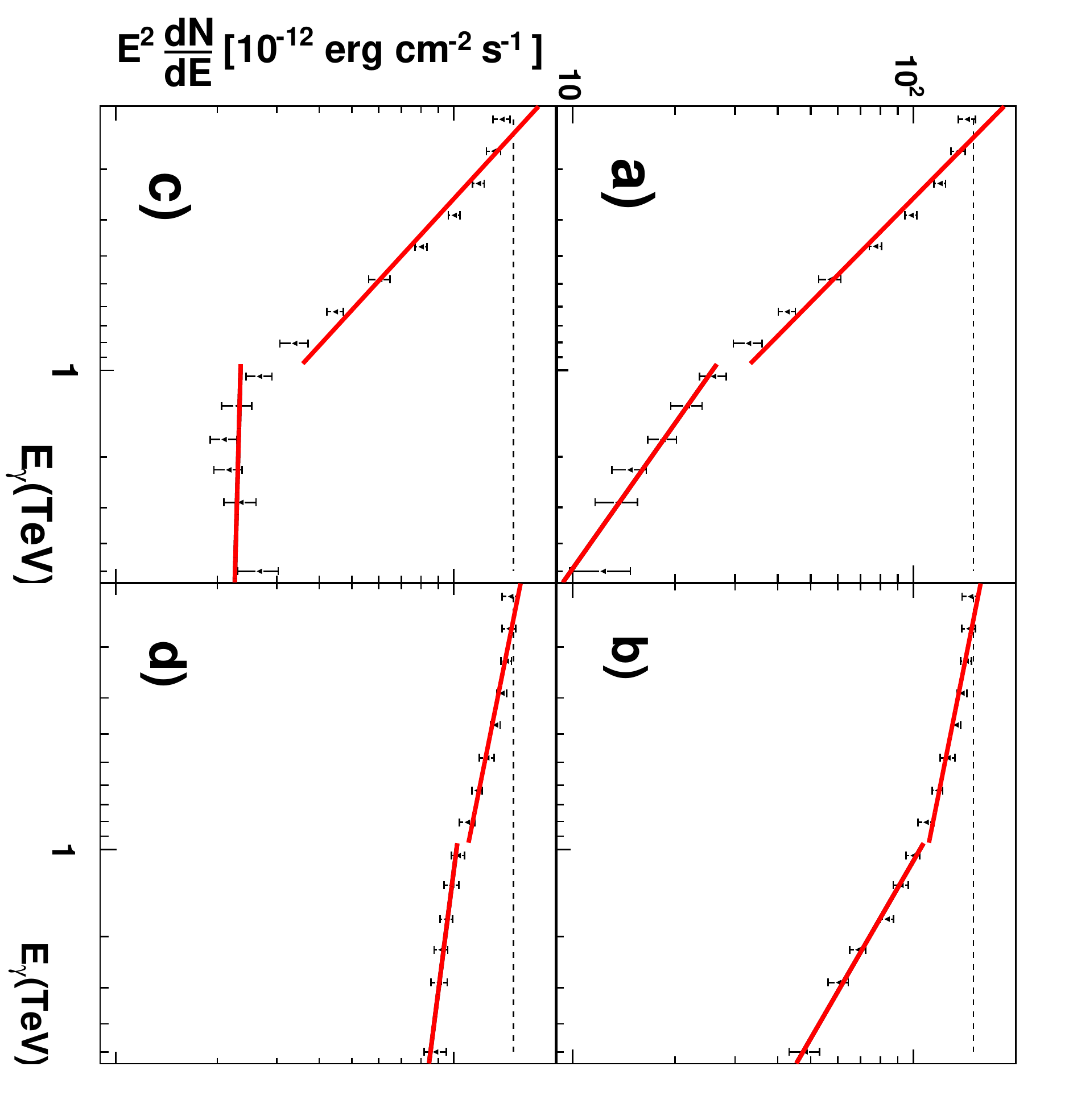}
\end{center}
\caption{absorbed energy spectra of a blazar for a source at  {\it z}=0.048 with a power law $\sim$ $\rm E^{-2}$ using different EBL scenarios: a) HHH, b) LHH, c) HLL and d) LLL. }\label{fig3}
\end{figure}
\begin{figure}
\begin{center}
\includegraphics [scale=0.75 , angle=90,width=0.48\textwidth]{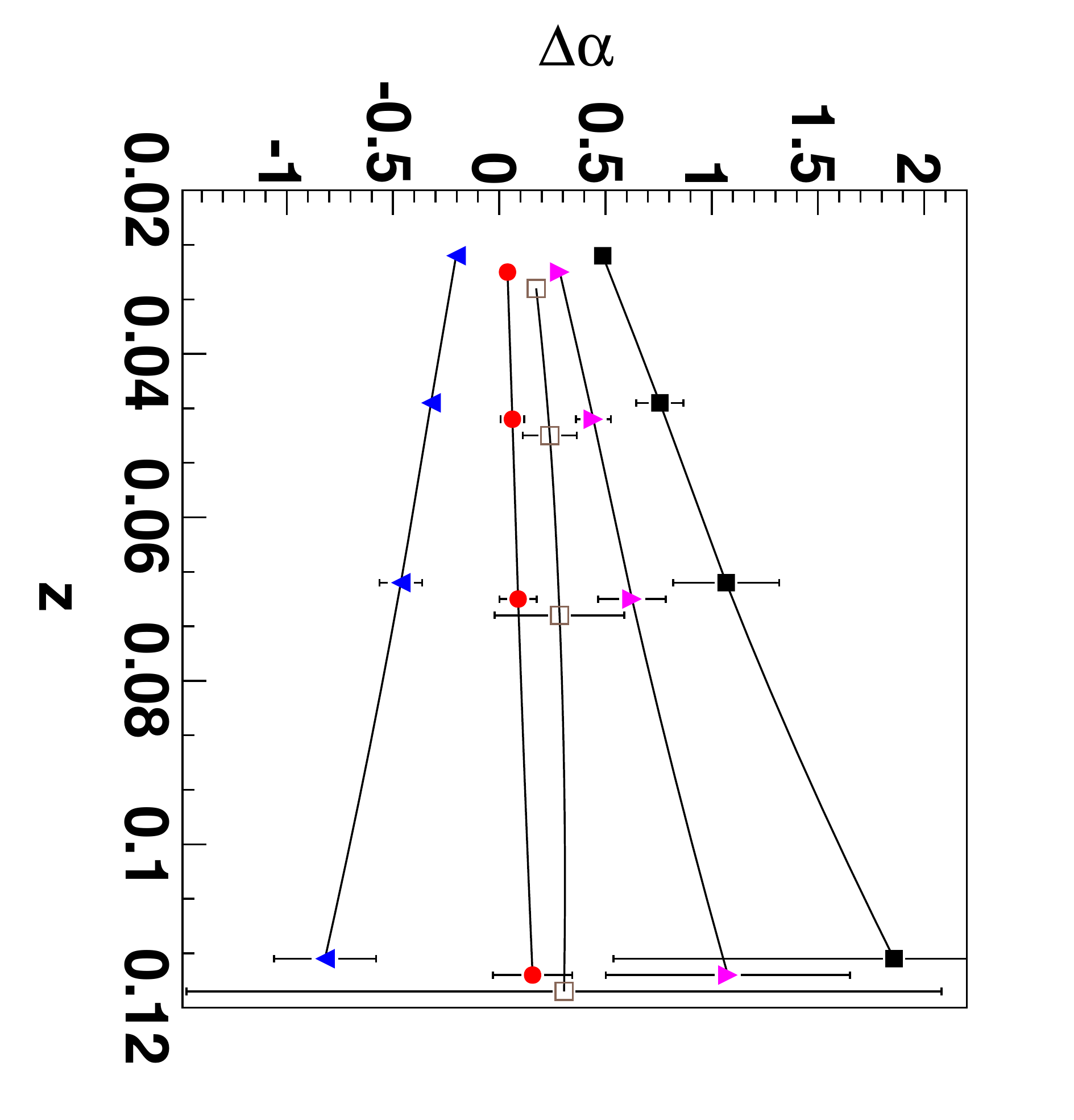}
\end{center}
\caption{the spectral break as a function of redshift is shown for different EBL scenarios. Shown here are
{\it solid squares}-HLL, {\it solid triangle}-MLL, {\it open squares}-HHH, {\it solid circles}-LLL, {\it solid inverted triangles}-LHH. Plotted are points at redshifts of {\it z} = 0.03, 0.048, 0.07, 0.116  only but shown with minor offsets for legibility.}\label{fig4}
\end{figure}

Furthermore, we assume power law source spectra since the EBL absorption of interest can be derived from the limited energy range between 0.1 and 5~TeV\footnote{Strongly curved intrinsic spectra or cutoffs are less sensitive for searching for a spectral feature.}. Figure 3  shows attenuation by four different EBL scenarios of a theoretical $\rm E^{-2} $ source at z=0.048. 
The modulation of the spectral indices at 1 TeV is  characterized by simple power law fits, $\rm E^{-\alpha}$,  providing $\alpha_{< 1 TeV}$ and above 1 TeV region $\alpha_{> 1 TeV}$. The change in the spectral index is quantified by  $\Delta \alpha$, the difference between $\alpha_{< 1 TeV}$ and  $\alpha_{> 1 TeV}$.
 We clearly see that the magnitude of $\Delta \alpha$ is mediated by the relative values of the EBL photon densities in the near, mid and far IR.  Curved source spectra or sources with intrinsic cutoffs provide diminished sensitivity for this analysis as they rapidly drop off.  The best sources for searching for a spectral signature are HBLs (High frequency selected BL Lacs) with peak energies in the multi-TeV regime. 
It is noteworthy, that the change in spectral index, $\Delta \alpha$, is essentially independent of the source spectral index. This feature could allow us to discriminate among competing EBL scenarios with out making assumptions about source models as this method merely relies on finding a spectral break at 1 TeV. Most importantly, the magnitude of $\Delta \alpha $ increases with redshift, see figure 4. Hence, sources at larger redshifts provide the best prospects for finding such a break.  With the given assumptions about the flaring levels the search for a spectral break is most promising at redshift of {\it z} $\sim$ 0.05 to 0.13. 
Based on our calculations, the absence of a noticable sprectral break in the nearby strongly flaring blazars can be jointly attributed to the fact that the expected magnitude of $\Delta \alpha $ is small and the spectra of these nearby blazars are intrinsically curved.

 %=======================================
\section{Conclusions}
 %=======================================
We have shown that with VERITAS/HESS exposure times of 50 hours could provide sufficient statistics
to identify an  EBL induced absorption feature in the TeV spectra of blazars. A systematic study of 
many HBL blazar candidates with peak energies in the multi-TeV regime are most promising to finding
an EBL signature, especially at redshifts of 0.05 - 0.13. 

 %=======================================
\section{Acknowledgements}
 %=======================================
F.K. acknowledges support by the US Department of Energy.
%This is the reference to .bib file (Whitout .bib!)
\bibliography{icrc0683}
%This in the bibtex style, is ok.
\bibliographystyle{plain}

\end{document}